\documentclass[runningheads]{llncs}
\usepackage{graphicx}
\usepackage[utf8]{inputenc}

\usepackage{physics} 
\usepackage{hyperref}
\usepackage{amsmath,amsfonts,amssymb} 
\usepackage{xcolor}
\usepackage{tabularx}
\usepackage{amsfonts}
\usepackage{array,collcell}
\usepackage{comment}
\usepackage{pifont}
\usepackage{multirow}

\newcommand\AddLabel[1]{%
  \refstepcounter{equation}
  (\theequation)
  \label{#1}
}
\newcolumntype{M}{>{$\displaystyle}X<{$\hfil}} 
\newcolumntype{L}{>{\collectcell\AddLabel}r<{\endcollectcell}}
\newcolumntype{P}[1]{>{\centering\arraybackslash}p{#1}}

\begin{document}

\title{Celestial Machine Learning}
\subtitle{Discovering the Planarity, Heliocentricity, and Orbital Equation of Mars with AI Feynman}

%
%
\author{Zi-Yu Khoo\inst{1}, Gokul Rajiv\inst{1}, Abel Yang\inst{1}, Jonathan Sze Choong Low\inst{2}, \\ and
Stéphane Bressan\inst{1}\inst{3} }
\authorrunning{Khoo et al.}

\institute{National University of Singapore. 21 Lower Kent Ridge Rd, Singapore 119077
\email{khoozy@comp.nus.edu.sg, grajiv@u.nus.edu, phyyja@nus.edu.sg, steph@nus.edu.sg}\\
\and Singapore Institute of Manufacturing Technology, Agency for Science, Technology and Research (A*STAR), Singapore 138634 \email{sclow@simtech.a-star.edu.sg}
\and CNRS@CREATE LTD, 1 Create Way, Singapore 138602\\
}
\maketitle              
\begin{abstract}
Can a machine or algorithm discover or learn the elliptical orbit of Mars from astronomical sightings alone? 
Johannes Kepler required two paradigm shifts to discover his First Law regarding the elliptical orbit of Mars. Firstly, a shift from the geocentric to the heliocentric frame of reference. Secondly, the reduction of the orbit of Mars from a three- to a two-dimensional space. We extend AI Feynman, a physics-inspired tool for symbolic regression, to discover the heliocentricity and planarity of Mars' orbit and emulate his discovery of Kepler's first law.

\end{abstract}
\section{Introduction}
In 2020, Silviu-Marian Udrescu and Max Tegmark introduced \emph{AI Feynman}~\cite{Udrescueaay2631}, a symbolic regression algorithm that could rediscover one hundred equations from the \emph{Feynman Lectures on Physics}~\cite{feynmanlecture}. The authors motivated their work with Johannes Kepler's rediscovery of the orbital equation of Mars. Recent work has shown that AI Feynman can also emulate Johannes Kepler's rediscovery of the orbital equation of Mars from the Rudolphine tables~\cite{rudolphine_tables} when AI Feynman is biased with information regarding the periodicity of Mars’ orbit, and the trigonometric nature of the data from the Rudolphine tables~\cite{khoo_2023}. 


In this work, we again use AI Feynman to emulate Kepler's rediscovery of the elliptical orbital equation of Mars, but from astronomical observations instead of tabulated data from the Rudolphine tables. Kepler's rediscovery of the orbital equation of Mars required two paradigm shifts in early astronomy which he embedded within the Rudolphine tables. Firstly, a shift from the geocentric to the heliocentric frame of reference. Secondly, the reduction of the orbit of Mars from a three-dimensional space to a two-dimensional plane. Without observations from the Rudolphine tables, AI Feynman is ignorant of these two paradigm shifts and unable to emulate Kepler's rediscovery of the orbital equation of Mars. 

Our work extends AI Feynman to discover the heliocentricity and planarity of Mars' orbit. We embed the ability to change reference frames and reduce dimensions in AI Feynman via biases in the style of physics-informed machine learning. Embedding these biases in AI Feynman allows it to discover relationships otherwise obscured by the perspective of the observer or concealed in higher dimensional data. We devise, present, and evaluate the performance of two algorithms that allow AI Feynman to rediscover the orbital equation of Mars by smartly exploring changes in reference frames and reductions in dimension spaces.

Our approach imbues AI Feynman with a level of understanding akin to Kepler's paradigm-shifting insights, facilitating the discovery of complex physical relationships concealed within the observer's perspective or hidden in higher-dimensional data. The remainder of this paper details this significant advancement in symbolic regression algorithms and evaluates its performance.


\section{Background} ~\label{sec:background}
Kepler's First Law states that each planet's orbit about the Sun is an ellipse with the Sun's center located at one focus of the ellipse. The planet follows the ellipse in its orbit, and the planet to Sun distance is constantly changing as the planet goes around its orbit~\cite{kepler_law_nasa}. The orbital equation of Mars in polar coordinates is shown in Equation~\ref{eqn:mars_polar}. It describes the distance of Mars from the Sun, $r$, as a function of the \textit{anomalia coaequata}, $\theta$, an angle between Mars and the Sun with respect to a horizontal. $a$ and $\epsilon$ are constants representing the semi-major axis and eccentricity of the ellipse. The variables $r(t)$ and $\theta(t)$ are functions of time, and represent vectors comprising observations of that variable at multiple different times.
\begin{equation}
    r(t) = \frac{a}{1+\epsilon \cos(\theta(t))} \label{eqn:mars_polar} 
\end{equation}
Equation~\ref{eqn:mars_polar} describes Mars' orbit as an ellipse centered on its focus. An alternative representation of an ellipse centered on its focus is shown in Equation~\ref{eqn:ellipse_cartesian}. $x$ and $y$ are the coordinates of Mars relative to the Sun, and $a$, $b$ and $h$ are the semi-major axis, semi-minor axis and distance between the center and focus of the ellipse respectively.

\begin{equation}
    y(t)^2 = b^2 \times \left(1-\frac{(x(t)-h)^2}{a^2} \right) \label{eqn:ellipse_cartesian} 
\end{equation}
The distance $r(t)$ between Mars and the Sun in cartesian coordinates is shown in Equation~\ref{eqn:mars_cartesian}. It is a modification of Equation~\ref{eqn:ellipse_cartesian}.
\begin{equation}
    r(t) = \sqrt{y(t)^2+x(t)^2} = \sqrt{b^2 \times \left(1-\frac{(x(t)-h)^2}{a^2} \right)+x(t)^2} \label{eqn:mars_cartesian} 
\end{equation}

Equations~\ref{eqn:mars_polar}, \ref{eqn:ellipse_cartesian} and \ref{eqn:mars_cartesian} are equivalent in that they either directly or indirectly describe Mars' orbit around the Sun.

\section{Related Work} ~\label{sec:relatedwork}
\subsection{Symbolic Regression}
Finding an equation from sample data, for instance, an equation describing the orbit of Mars from sightings of the planet and the Sun, is a combinatorial challenge~\cite{Udrescueaay2631}. To circumvent this, one may use universal function approximators such as multilayer perceptron neural networks~\cite{Hornik_Stinchcombe_White_1989}. 

Alternatively, symbolic regressions search for a parsimonious and elegant form of the unknown equation. There are three main classes of symbolic regression methods~\cite{Makke:2022rnq}: regression-based, expression tree-based and physics- or mathematics-inspired. We use AI Feynman, a physics-inspired algorithm~\cite{Udrescueaay2631}. 

Regression-based symbolic regression methods~\cite{Makke:2022rnq}, given solutions to the unknown equation, find the coefficients of a fixed basis that minimise the prediction error. As the basis grows, the fit improves, but the functional form of the unknown equation becomes less sparse or parsimonious. Sparse regressions promote sparsity through regularisation, as proposed by Robert Tibshirani~\cite{Tibshirani94regressionshrinkage} who used the $\mathit{l}_1$ norm, thus inventing the Lasso regression. Steven Brunton et al.'s Sparse Identification of Nonlinear Dynamics~\cite{Brunton3932} is a state-of-the-art sparse symbolic regression approach. It leverages regularisation and identifies a system's equations of motion using a sparse regression over a chosen basis. However, committing to a basis limits the applicability of regression-based methods. 

Expression tree-based symbolic regression methods based on genetic programming~\cite{Makke:2022rnq} can instead discover the form and coefficients of the unknown equation.  The seminal work by John Koza~\cite{Koza92} represents each approximation of an unknown equation as a genetic programme with a tree-like data structure, with traits (or nodes in the tree) representing functions or operations and variables representing real numbers. The fitness of each genetic programme is its prediction error. Fitter genetic programmes undergo a set of transition rules comprising selection, crossover and mutation to find the optimal equation form iteratively. Genetic programmes may greedily mimic nuances of the unknown equation~\cite{Smits2005}, limiting generalisability. 
David Goldberg~\cite{Goldberg1989} used Pareto optimisation to balance the objectives of fit and parsimony in symbolic regression. 
State-of-the-art symbolic regression using genetic programming includes \verb|Eureqa| by Michael Schmidt and Hod Lipson~\cite{Schmidt81} and \verb|PySR| by Miles Cranmer~\cite{pysr}. 
However, if an expression tree-based method finds a reasonably accurate equation with the wrong functional form, it risks getting stuck at a local optimum~\cite{Udrescueaay2631}. 

Physics-inspired symbolic regression methods leverage properties of the unknown equation like symmetry and separability~\cite{Udrescueaay2631}. Udrescu and Tegmark~\cite{Udrescueaay2631}, in AI Feynman, use a neural network to test for such properties and recursively break the search for the unknown equation into that of simpler equations~\cite{Udrescueaay2631}. Each equation is then regressed with a basis-set of nonlinear functions. This guarantees that more accurate approximations of an equation are symbolically closer to the truth~\cite{Udrescueaay2631}. AI Feynman outputs a sequence of increasingly complex equations with progressively better accuracy along a Pareto frontier, leveraging the work of Goldberg~\cite{Goldberg1989} and Smits~\cite{Smits2005} to balance fit and parsimony. We use AI Feynman to rediscover the orbital equation of Mars from observational data.

\subsection{Physics-Informed Machine Learning}
Karniadakis et al. focus on three modes of biasing a learning algorithm: observational bias, learning bias, and inductive bias~\cite{Karniadakis2021}. 
Observational biases can be introduced directly through data that embody the underlying physics or carefully crafted data augmentation procedures. With sufficient data to cover the input domain of a learning task, machine learning methods have demonstrated remarkable power in achieving accurate interpolation between the dots~\cite{Karniadakis2021}. 
Learning biases can be introduced by appropriate loss functions, constraints and inference algorithms that modulate the training phase of a machine learning model to explicitly favour convergence towards solutions that adhere to the underlying physics~\cite{Karniadakis2021}. These are soft constraints~\cite{Karniadakis2021}. 
Inductive biases are prior assumptions incorporated by tailored interventions to a machine learning model architecture, so predictions are guaranteed to implicitly satisfy a set of given physical laws~\cite{Karniadakis2021}. These are strict constraints. We embed biases within the AI Feynman algorithm so that it can emulate Kepler's change in reference frames, to rediscover the orbital equation of Mars.

Seminal work by Khoo et al.~\cite{khoo_2023} combined AI Feynman and physics-informed machine learning to rediscover the orbital equation of Mars. However, this work made use of data from the Rudolphine tables that already embeded assumptions of the heliocentricity and planarity of Mars' orbit~\cite{khoo_2023}. Geocentric observations of Mars and the Sun do not embed any assumptions. Therefore, in rediscovering the orbital equation of Mars from geocentric observations, AI Feynman needs to be informed of these assumptions of heliocentricity and planarity of Mars' orbit. Our work focuses on embedding biases regarding the heliocentricity and planarity of Mars' orbit within the AI Feynman algorithm. 

Additionally, the authors of the previous work needed to embed additional inductive and observational biases within the AI Feynman algorithm to inform the algorithm of the periodicity of Mars' orbit and the trigonometric nature of the data from the Rudolphine tables via two biases~\cite{khoo_2023}. The first was an observational bias which replaced angular values with their sine and cosine. The second was an inductive bias which restricted the search space. The bias stemmed from the knowledge that exponential and logarithmic functions only transform dimensionless quantities, therefore cannot transform data representing physical quantities and limited the search space to trigonometric, polynomial and radical functions~\cite{khoo_2023}. The necessity of embedding these biases arose from the data in the Rudolphine tables being presented in polar coordinates. AI Feynman, augmented with the observational and inductive biases to inform the algorithm of the periodicity of Mars' orbit and the trigonometric nature of the data, was best able to rediscover Mars' orbital equation. 

\section{Methodology} \label{sec:methodology}
This section introduces two algorithms that correspond to two biases to aid AI Feynman in rediscovering the orbital equation of Mars. The first algorithm considers the relationship between Mars and the Sun from different reference frames, namely the heliocentric, geocentric and areocentric (centered around Mars) reference frames. The second considers the relationship between Mars and the Sun from varying dimensional spaces. We evaluate the two algorithms on their ability to rediscover either Equation~\ref{eqn:mars_polar}, \ref{eqn:ellipse_cartesian} or \ref{eqn:mars_cartesian}. 

\subsubsection{The First Algorithm}
The first algorithm considers the relationship between Mars and the Sun from varying reference frames. Observations of Mars, the Sun and the Earth are made from the three heliocentric, geocentric and areocentric reference frames. 

In the absence of information regarding the choice of reference frames, AI Feynman is used to rediscover the relationship between Mars and the Sun from each set of observations. A Pareto frontier comparing fit and parsimony is produced from each set of observations and the three Pareto fronts are combined to produce a single Pareto front. The combination of the three Pareto fronts is an inductive bias as equations along the single combined Pareto frontier implicitly optimise fit and parsimony. 

With information regarding the choice of reference frame, AI Feynman is used to rediscover the relationship between Mars and the Sun from observations of the selected reference frame. This is an observational bias where the choice of reference frame is introduced directly through observations that embody the underlying physics of the relationship between Mars and the Sun. 

\subsubsection{The Second Algorithm}
The second algorithm considers the relationship between Mars and the Sun from varying dimensional spaces. Three-dimensional geocentric observations of Mars and the Sun are made. 

In the absence of information regarding the choice of dimensionality, the dimensional space of geocentric observations of Mars and the Sun is gradually reduced by projecting the original observations into progressively lower dimensional spaces. AI Feynman is used to rediscover the relationship between Mars and the Sun from each set of lower-dimensional projections. A Pareto frontier comparing fit and parsimony is produced from each set of projections and the Pareto fronts are combined to produce a single Pareto front. The combination of the Pareto fronts is an inductive bias as equations along the single combined Pareto frontier implicitly optimise fit and parsimony. 

With information regarding the choice of dimensionality, AI Feynman is used to rediscover the relationship between Mars and the Sun from projections in the selected dimensional space. This is an observational bias where the choice of dimensionality is introduced directly through observations that embody the underlying physics of the relationship between Mars and the Sun. 

We evaluate the algorithms on their ability to rediscover equations that describe Mars' orbit around the Sun, namely Equations~\ref{eqn:mars_polar} \ref{eqn:ellipse_cartesian} or \ref{eqn:mars_cartesian}, from observations from a polar or cartesian coordinate system. 

\section{Performance Evaluation} \label{sec:exp}
Three experiments are designed. The first two correspond to the two algorithms described in Section~\ref{sec:methodology} in the absence of information regarding the choice of reference frame and dimensionality. Their experimental setup and results are shown in subsections~\ref{sec:exp1} and \ref{sec:exp2}. The third experiment, in subsection~\ref{sec:exp3} is a proof of concept that informs AI Feynman via observational biases about the choice of reference frame and dimensionality to rediscover the orbital equation of Mars. All experiments use different inputs, elaborated on in subsections~\ref{sec:exp1}, \ref{sec:exp2} and \ref{sec:exp3}, to describe the distance between Mars and the Sun. All experiments are run twice, once each for the cartesian and polar coordinate systems respectively.

Data for all experiments was obtained from the National Aeronautics and Space Administration's Horizons system~\cite{horizon_nasa} and downloaded from \verb|astropy|~\cite{astropy:2022}. The data included geocentric observations of the angular width, $AW$, right ascension, $RA$ and declination, $DC$, of Mars $M$ and the Sun $S$ from 1 January 1601 to 31 December 1602. All computations are in astronomical units. The code and data are available at \verb|https://github.com/zykhoo/AI-Feynman|.

From the angular width, the distances $DT$ of Mars and the Sun from the Earth are computed following Equations~\ref{eqn:dist_M} and \ref{eqn:dist_S}. $SF_{AU}$ is a scaling factor representing the diameter of the Sun relative to one astronomical unit, with a value of $0.00465047$. $SF_M=0.00486759$ and $SF_S=1$ are scaling factors representing the diameter of Mars relative to the Sun and the diameter of the Sun relative to itself respectively. The angular width, right ascension and distance are polar coordinates. The geocentric cartesian coordinates of Mars is a tuple $(X_M, Y_M, Z_M)$, converted from polar coordinates using \verb|astropy|.\begin{equation}
    DT_M(t) = \frac{SF_M \times SF_{AU}}{\tan(AW_M(t)/2)} \label{eqn:dist_M}
\end{equation}
\begin{equation}
    DT_S(t) = \frac{SF_S \times SF_{AU}}{\tan(AW_S(t)/2)} \label{eqn:dist_S}
\end{equation}

The AI Feynman algorithm designed by Udrescu et al.~\cite{Udrescueaay2631} with modifications to embed inductive and observational biases by Khoo et al.~\cite{khoo_2023} is used. The modifications to embed an inductive bias inform AI Feynman of the periodicity of Mars' orbit and the trigonometric nature of its input data by the restriction of the search space. This inference stems from the knowledge that exponential and logarithmic functions only transform dimensionless quantities, therefore cannot transform data representing physical quantities. The inductive bias limits the search space to trigonometric, polynomial and radical functions. The modifications to embed an observational bias inform AI Feynman of the periodicity of Mars' orbit and the trigonometric nature of its input data replacing angular values with their sine and cosine.

AI Feynman returns equations along a Pareto frontier. The Pareto frontier is a set of solutions that represents the best trade-off between fit and parsimony. No other solution has a better fit and is more parsimonious than an equation along the Pareto frontier. The fit and parsimony of equations along the Pareto frontier are optimised using the minimum description length formalism~\cite{tailin2019_mdl,wutailin_thesis}. Minimising this description length formalism is equivalent to minimising the number of bits required to represent the equation and its loss. The minimum description length formalism comprises a measure of fit and a measure of parsimony. The measure of fit places a logarithm-scaled penalty on the absolute loss. The measure of parsimony places a logarithm-scaled penalty on real numbers, variables and operators in an equation. AI Feynman simultaneously attempts to minimise both penalties and this minimises the geometric mean instead of the arithmetic mean, which encourages improving already well-fit points~\cite{tailin2019_mdl,wutailin_thesis}.

\subsection{Experiment 1} \label{sec:exp1}
This subsection details the experimental setup and implementation of the first algorithm, and its results in rediscovering the relationship between Mars and the Sun from heliocentric, geocentric and areocentric reference frames. 

\subsubsection{Experimental Setup}
Three sets of observations are created, corresponding to the three reference frames for each coordinate system. The first set comprises the coordinates of Mars and the Sun from the geocentric reference frame, the second set comprises the coordinates of the Earth and Mars from the heliocentric reference frame, and the third set comprises the coordinates of the Sun and the Earth from the areocentric reference frame. The heliocentric and areocentric reference frames of Mars, the Earth and the Sun are computed using vector addition of the coordinates in the geocentric reference frame. 

For the cartesian coordinate system, the x, y and z coordinates of the two bodies in each of the three reference frames are input to AI Feynman to describe the distance between Mars and the Sun. Independent runs of AI Feynman are repeated for the three reference frames. The fit and parsimony of equations output by AI Feynman from the three runs are compared on the same Pareto frontier. The equations that form the new, combined Pareto frontier are examined.

For the polar coordinate system, the declination and right ascension are replaced with their sines and cosines as an observational bias regarding the periodicity of Mars' orbital equation. The sine and cosine of the declination, the sine and cosine of the right ascension and the angular width are input to AI Feynman to describe the distance between Mars and the Sun. Independent runs of AI Feynman are repeated for the three reference frames. The fit and parsimony of equations output by AI Feynman from the three runs are compared on the same Pareto frontier and the equations that form the new Pareto frontier are examined.

\subsubsection{Experimental Results} \label{sec:exp1_results}
The equations along the new, combined Pareto frontiers are reported in column 2 of Tables~\ref{tab:exp1_results} and \ref{tab:exp1_results_polar} for the cartesian and polar coordinate systems respectively. $r(t)$ denotes the distance between Mars and the Sun. $MS(t)$ and $ME(t)$ denote areocentric observations of the Sun and Earth, $SE(t)$ and $SM(t)$ denote heliocentric observations of Earth and Mars, and $EM(t)$ and $ES(t)$ denote geocentric observations of Mars and the Sun. The subscripts $x$, $y$ and $z$ denote the cartesian coordinate system, while the subscripts $RA$, $DC$ and $AW$ denote the right ascension, declination and angular width in the polar coordinate system. The fit and parsimony of the equations are also measured and presented in columns 3 and 4. In Table~\ref{tab:exp1_results}, Equation \ref{a1} is areocentric. Equations \ref{a3} to \ref{a9} are heliocentric. Equation \ref{a2} is geocentric. In Table~\ref{tab:exp1_results_polar}, Equation \ref{b4} is heliocentric. Equations \ref{b1}, \ref{b2} and \ref{b3} are geocentric. 

\begingroup
\begin{subequations}
\begin{table}[h]
    \tiny
    \centering
    \renewcommand{\arraystretch}{1.3} 
    \begin{tabularx}\textwidth{|@{}|L|M|c|c|@{}|}
    \hline
    \multicolumn{1}{|c|}{No.} & \multicolumn{1}{|c|}{Equation} & \multicolumn{1}{|c|}{Fit} & \multicolumn{1}{|c|}{Parsimony} \\ 
    \hline
        a1 & r(t) = \arccos(-0.1\times MS_x(t)) &  25.05 &  12.32  \\
        a2 & r(t) = -EM_y(t) + EM_z(t)\times ES_y(t)/ES_z(t) + 1.5 &  24.10 &  29.0\\
        a3 & r(t) = (SM_x(t)^2 + 1.17801253496029\times SM_y(t)^2)^{0.5} &  22.82 &  63.52\\
        a4 & \begin{aligned}
            r(t) &= (SM_x(t)^2 + 1.17801253496029\times SM_y(t)^2 \\
            \quad & + 0.0538855469457899)^{0.5}
        \end{aligned} &  22.61 &  105.82\\
        a5 & \begin{aligned}r(t) &= \arccos\left(-0.294567985407943\times SM_x(t)^2 \right.\\
            \quad & \left.- 0.360901709329182\times SM_y(t)^2 + 0.75\right)\end{aligned} &   22.52 &  112.46\\
        a6 & \begin{aligned}r(t) &= 0.346480692154936\times SM_x(t)^2 \\
            \quad & + 0.424521090029716\times SM_y(t)^2 + 0.705090417032556\end{aligned} &  21.99 &  151.03\\
        a7 & \begin{aligned}r(t) &= (0.978840855738558\times SM_x(t)^2 \\
            \quad & - 0.0297412516501264\times SM_x(t)\times SM_y(t) \\
            \quad & + 1.17801253496029\times SM_y(t)^2 + 0.0538855469457899)^{0.5}\end{aligned} &  21.23 &  197.63\\
        a8 & \begin{aligned}r(t) &= (0.978840855738558\times SM_x(t)^2 \\
            \quad & - 0.0297412516501264\times SM_x(t)\times SM_y(t) \\
            \quad & + 1.17801253496029\times SM_y(t)^2 + \\
            \quad & 0.00651713753890506\times SM_y(t) + 0.0538855469457899)^{0.5}\end{aligned} &   20.31 &  236.87\\
        a9 & \begin{aligned}r(t) &= (0.978840855738558\times SM_x(t)^2 \\
            \quad & - 0.0297412516501264\times SM_x(t)\times SM_y(t) \\
            \quad & - 0.00278630946992131\times SM_x(t) \\
            \quad &+ 1.17801253496029\times SM_y(t)^2 \\
            \quad &+ 0.00651713753890506\times SM_y(t) + 0.0538855469457899)^{0.5}\end{aligned} &  20.21 &  274.89\\
    \hline
    \end{tabularx}
    \caption{Results for Experiment 1 with cartesian coordinates}
    \label{tab:exp1_results}
\end{table}
\end{subequations}
\endgroup        

\begingroup
\begin{subequations}
\begin{table}[h]
    \tiny
    \centering
    \renewcommand{\arraystretch}{1.3} 
    \begin{tabularx}\textwidth{|@{}|L|M|c|c|@{}|}
    \hline
    \multicolumn{1}{|c|}{No.} & \multicolumn{1}{|c|}{Equation} & \multicolumn{1}{|c|}{Fit} & \multicolumn{1}{|c|}{Parsimony} \\ 
    \hline
        b1 & r(t) = 2\times \cos(EM_{DC}(t)) - 0.4 &  25.92 &  10.25 \\
        b2 & r(t) = 1.66666666666667\times \cos(EM_{DC}(t))^2 &  25.57 &  10.51\\
        b3 & r(t) = 6\times (\cos(EM_{DC}(t))\times EM_{AW}(t)^{0.25})^{0.5} &  25.35 &  15.81 \\
        b4 & r(t) = 0.000045273194\times((0+1)/MS_{AW}(t)) &  3.25 &  35.08 \\
    \hline
    \end{tabularx}
    \caption{Results for Experiment 1 with polar coordinates}
    \label{tab:exp1_results_polar}
\end{table}
\end{subequations}
\endgroup   

        


We observe that Equations \ref{a3}, \ref{a7}, \ref{a8} and \ref{a9} from Table~\ref{tab:exp1_results} are similar to the equation form for Equation~\ref{eqn:mars_cartesian}. From Table~\ref{tab:exp1_results_polar}, we can observe that no equation matches the form of Equation~\ref{eqn:mars_polar}, although Equation \ref{b4} in Table~\ref{tab:exp1_results_polar} is an attempt to use angular width to describe the relationship between Mars and the Sun as seen in Equation~\ref{eqn:dist_M}. 
Furthermore, we can observe that most equations from both tables are heliocentric and use observations of Mars to describe $r(t)$. These equations suggest a relationship between Mars and the Sun that is independent of observations of the Earth. Such equations fit the data parsimoniously, therefore Pareto-dominate other equations that use observations of the Earth which have a poorer fit or are less parsimonious. Therefore, heliocentric equations appear on the combined Pareto front. 

Lastly, we note the absence of heliocentric equations along the combined Pareto frontier presented in Table~\ref{tab:exp1_results_polar}. This may be because the heliocentric equations are obscured by the high dimensional data from the areocentric and geocentric reference planes. However, one equation of interest along the heliocentric Pareto frontier is $r(t) = 0.009300823088\times((0+1)/SM_{AW})$, which is heliocentric and has the same form as Equation \ref{b4}. It is an attempt to use angular width to describe the relationship between Mars and the Sun as seen in Equation~\ref{eqn:dist_S} from a heliocentric frame of reference. This equation does not appear along the combined Pareto frontier as it is Pareto dominated by Equation \ref{b4}.

\subsection{Experiment 2} \label{sec:exp2}
This subsection details the experimental setup and results in the implementation of the second algorithm, which rediscovers the relationship between Mars and the Sun from varying dimensional spaces.

\subsubsection{Experimental Setup}
The three-dimensional geocentric observations of Mars and the Sun in both the polar and cartesian coordinate systems are reduced to lower dimensional spaces by principal component analysis~\cite{Jolliffe1986}. 

For the cartesian coordinate system, the three-dimensional geocentric x, y and z coordinates of Mars and the Sun are projected to three-dimensional, two-dimensional and one-dimensional spaces. The three-dimensional geocentric x, y and z coordinates of Mars, $M(t)$, and the Sun, $S(t)$, are expressed as $M(t) = [M_x(t), M_y(t), M_z(t)]$ and $S(t) = [S_x(t), S_y(t), S_z(t)]$ respectively. The covariance matrices for Mars and the Sun are $M(t)^\intercal M(t)$ and $S(t)^\intercal S(t)$ respectively. We skip standardisation to retain measurements in astronomical units. The eigenvalues and eigenvectors of the respective covariance matrices are computed. To get a lower-dimensional projection of the three-dimensional space, $M(t)$ is projected onto an N-dimensional space defined by the eigenvectors corresponding to the N largest eigenvalues of $M(t)^\intercal M(t)$ and $S(t)$ is projected onto an N-dimensional space defined by the eigenvectors corresponding to the N largest eigenvalues of $S(t)^\intercal S(t)$. This is repeated for $N = [3, 2, 1]$. 

Three independent runs of AI Feynman are made. The inputs to each run are the six projections of Mars and the Sun on to a three-dimensional space, four projections of Mars and the Sun on to a two-dimensional space and two projections of Mars and the Sun on to a one-dimensional space respectively. All runs attempt to describe the distance between Mars and the Sun. The fit and parsimony of equations output by AI Feynman from the three runs are compared on a combined Pareto frontier and the equations that form the new, combined Pareto frontier are examined.

For polar coordinates, the three-dimensional polar coordinates of Mars and the Sun comprise the declination, right ascension and angular width. As the declination and right ascension are angles and the angular width is a measure of length, they cannot be simultaneously projected onto a lower dimensional space. Therefore, 
the angles have to be projected onto a lower-dimensional space, after which they are replaced with their sines and cosines as an observational bias regarding the periodicity of Mars' orbital equation~\cite{khoo_2023}. 

For the projection of the angles onto a lower dimensional space, the angles of Mars and the Sun are expressed as $M(t) = [M_{RA}(t), M_{DC}(t)]$ and $S(t) = [S_{RA}(t), S_{DC}(t)]$ respectively. The covariance matrices for Mars and the Sun are $M(t)^\intercal M(t)$ and $S(t)^\intercal S(t)$ respectively. The eigenvalues and eigenvectors of the covariance matrices are computed. To get an N-dimensional projection of the two-dimensional space, $M(t)$ is projected onto an N-dimensional space defined by the eigenvectors corresponding to the N largest eigenvalues of $M(t)^\intercal M(t)$ and $S(t)$ is projected onto an N-dimensional space defined by the eigenvectors corresponding to the N largest eigenvalues of $S(t)^\intercal S(t)$, for $N = [2, 1]$.


Two independent runs of AI Feynman are made. 
The inputs to each run are the projections of the angles onto a lower dimensional space, after which they are replaced with their sines and cosines. The first run uses the sine and cosine of the projections in a two-dimensional space, and the angular width, for both the Sun and Mars. The second run uses the sine and cosine of the projections in a one-dimensional space, and the angular width, for both the Sun and Mars. All runs attempt to describe the distance between Mars and the Sun.
The fit and parsimony of equations output by AI Feynman from the two runs are compared on the same Pareto frontier and the equations that form the new, combined Pareto frontier are examined.

\subsubsection{Experimental Results}
The equations along the new, combined Pareto frontiers are reported in column 2 of Tables~\ref{tab:exp2_results} and \ref{tab:exp2_results_polar} for the cartesian and polar coordinate systems respectively. $\lambda_{M,N}(t)$ and $\lambda_{S,N}(t)$ denote the projection of geocentric observations of Mars and the Sun respectively, with $\lambda_{M,1}(t)$ and $\lambda_{S,1}(t)$ corresponding to the projections from the respective eigenvectors with the largest eigenvalues. For the cartesian and polar coordinate systems, $N \leq 3$ and $N \leq 2$ respectively. The fit and parsimony of the equations are also measured and presented in columns 3 and 4. In Table~\ref{tab:exp2_results}, Equation \ref{c6} is the output from a one-dimensional space. Equations \ref{c3}, \ref{c8}, \ref{c9}, \ref{c10}, \ref{c11}, \ref{c14}, \ref{c15} and Equations \ref{c1}, \ref{c2}, \ref{c4}, \ref{c5}, \ref{c7}, \ref{c12}, \ref{c13} are the outputs from two-dimensional and three-dimensional spaces respectively. In Table~\ref{tab:exp2_results_polar}, Equations \ref{d1}, \ref{d2} and \ref{d3} are outputs from a one-dimensional space. Equation \ref{d4} is the output from a two-dimensional space. 

None of the equations from Table~\ref{tab:exp2_results} and \ref{tab:exp2_results_polar} match the equation form for Equations~\ref{eqn:mars_cartesian} and \ref{eqn:mars_polar}. However, Equations~\ref{c14} and \ref{c15} make use of a square root to fit $r(t)$, similar to Equation~\ref{eqn:mars_cartesian}. Furthermore, even though the equations in Tables~\ref{tab:exp2_results} and \ref{tab:exp2_results_polar} are outputs from multi-dimensional spaces, most equations only make use of one to two eigenvectors from both Mars and the Sun. These equations suggest a relationship between Mars and the Sun that can be captured in a two-dimensional space. The eigenvectors from a planar, two-dimensional space fit the data parsimoniously, therefore Pareto-dominate other equations that make use of eigenvectors from higher dimensional spaces that have a poorer fit or are less parsimonious. Therefore, they appear on the combined Pareto front. 

\begingroup
\begin{subequations}
\begin{table}[h!]
    \tiny
    \centering
    \renewcommand{\arraystretch}{1.3} 
    \begin{tabularx}\textwidth{|@{}|L|M|c|c|@{}|}
    \hline
    \multicolumn{1}{|c|}{No.} & \multicolumn{1}{|c|}{Equation} & \multicolumn{1}{|c|}{Fit} & \multicolumn{1}{|c|}{Parsimony} \\ 
    \hline
        c1 & r(t) = 2\times \lambda_{M,3}(t)(t) + 1.5 &  26.17 &  9.34 \\
       c2& r(t) = 3\times \lambda_{M,3}(t)(t) + 1.5 &  26.13 &  9.75 \\
       c3& r(t) = \arccos(-0.1\times \lambda_{S,2}(t)) &  25.64 & 12.32 \\
       c4& r(t) = 2\times \lambda_{M,3}(t)(t) + 1.563343180129 &  25.62 & 53.49 \\
       c5& r(t) = 3\times \lambda_{M,3}(t)(t) + 1.554202964751 &  25.20 & 53.90 \\
       c6& r(t) = 1.221198383854+((0+1)/((((\lambda_{M,1}(t)+1)+1)+1)+1)) & 24.25 &  c54.87 \\
       c7& r(t) = 1.221364981489+((0+1)/((((\lambda_{M,1}(t)+1)+1)+1)+1)) & 24.22 &  54.87 \\
       c8& r(t) = 1.221468928502+((0+1)/((((\lambda_{M,1}(t)+1)+1)+1)+1)) & 24.19 &  54.87 \\
       c9& \begin{aligned}
           r(t) &= -0.0416098292506629\times \lambda_{M,1}(t)^2 - 0.0337764400695433\times \lambda_{M,1}(t)\\
           \quad & - 0.0526409552168828\times \lambda_{M,2}(t)^2 + 1.66666666666667
       \end{aligned} & 24.02 &  144.75 \\
       c10& \begin{aligned}
           r(t) &= (-0.017832447865237\times \lambda_{M,1}(t)^2 - 0.0122546126831841\times \lambda_{M,1}(t) \\
           \quad &- 0.0197336507795308\times \lambda_{M,2}(t)^2 \\
           \quad &- 0.00525510104504469\times \lambda_{M,2}(t) + 1.28929406165126)^2
       \end{aligned} & 22.89 &  223.87 \\
       c11& \begin{aligned}
           r(t) &= (-0.017832447865237\times \lambda_{M,1}(t)^2 - 0.000833616276921385\times \lambda_{M,1}(t)\times \lambda_{M,2}(t) \\
           \quad &- 0.0122546126831841\times \lambda_{M,1}(t) - 0.0197336507795308\times \lambda_{M,2}(t)^2 \\
           \quad &- 0.00525510104504469\times \lambda_{M,2}(t) + 1.28929406165126)^2
       \end{aligned} & 22.85 &  260.15 \\
       c12& \begin{aligned}
           r(t) &= \arccos\left(0.090616012160451\times \lambda_{M,1}(t) \right. \\
           \quad &- 0.0319600776983129\times \lambda_{M,2}(t) - 0.0232047508907192\times \lambda_{S,1}(t) \\
           \quad &+ 0.0283550675125619\times \lambda_{S,2}(t)^2 - 0.0976600194525864\times \lambda_{S,2}(t) \\
           \quad &+ 0.0289977987855869\times (\lambda_{M,3}(t)(t) - \left. \lambda_{S,1}(t) + \lambda_{S3}(t)\right)^2 + 0.0307948365787156)
       \end{aligned} &21.33 &  335.29 \\
       c13& \begin{aligned}
           r(t) &= \arccos\left(0.090616012160451\times \lambda_{M,1}(t) - 0.0319600776983129\times \lambda_{M,2}(t) \right. \\
           \quad &- 0.0232047508907192\times \lambda_{S,1}(t) + 0.0283550675125619\times \lambda_{S,2}(t)^2 \\
           \quad &- 0.0976600194525864\times \lambda_{S,2}(t) + 0.0232047508907192\times \lambda_{S3}(t) \\
           \quad &\left.+ 0.0289977987855869\times (\lambda_{M,3}(t)(t) - \lambda_{S,1}(t) + \lambda_{S3}(t))^2 + 0.0307948365787156\right)
       \end{aligned} & 21.32 &  376.36 \\
       c14& \begin{aligned}
           r(t) &= \left(0.618804692832026\times \lambda_{M,1}(t)^2 \right.- 0.63643067097383\times \lambda_{M,1}(t)\times \lambda_{S,1}(t) \\
           \quad &- 1.06306011300215\times \lambda_{M,1}(t)\times \lambda_{S,2}(t) - 0.104724154100595\times \lambda_{M,1}(t) \\
           \quad &+ 0.613759226153589\times \lambda_{M,2}(t)^2 - 1.05609058819238\times \lambda_{M,2}(t)\times \lambda_{S,1}(t) \\
           \quad &+ 0.632982631682707\times \lambda_{M,2}(t)\times \lambda_{S,2}(t) + 0.0355913865390322\times \lambda_{M,2}(t) \\
           \quad &+ 0.629510568181755\times \lambda_{S,1}(t)^2 + 0.0223685214839701\times \lambda_{S,1}(t) \\
           \quad &+ 0.612538055894269\times \lambda_{S,2}(t)^2 + 0.111425533671115\times \lambda_{S,2}(t) \\
           \quad &+ 0.871191849870558)^{0.5}
       \end{aligned} & 20.65 &  616.10 \\
       c15& \begin{aligned}
           r(t) &= (0.618804692832026\times \lambda_{M,1}(t)^2 - 0.63643067097383\times \lambda_{M,1}(t)\times \lambda_{S,1}(t) \\
           \quad &- 1.06306011300215\times \lambda_{M,1}(t)\times \lambda_{S,2}(t) - 0.104724154100595\times \lambda_{M,1}(t) \\
           \quad &+ 0.613759226153589\times \lambda_{M,2}(t)^2 - 1.05609058819238\times \lambda_{M,2}(t)\times \lambda_{S,1}(t) \\
           \quad &+ 0.632982631682707\times \lambda_{M,2}(t)\times \lambda_{S,2}(t) + 0.0355913865390322\times \lambda_{M,2}(t) \\
           \quad &+ 0.629510568181755\times \lambda_{S,1}(t)^2 + 0.00362394983019903\times \lambda_{S,1}(t)\times \lambda_{S,2}(t) \\
           \quad &+ 0.0223685214839701\times \lambda_{S,1}(t) + 0.612538055894269\times \lambda_{S,2}(t)^2 \\
           \quad &+ 0.111425533671115\times \lambda_{S,2}(t) + 0.871191849870558)^{0.5}
       \end{aligned} & 20.65 &  654.50 \\
    \hline
    \end{tabularx}
    \caption{Results for Experiment 2 with cartesian coordinates}
    \label{tab:exp2_results}
\end{table}
\end{subequations}
\endgroup   

\begingroup
\begin{subequations}
\begin{table}[h]
    \tiny
    \centering
    \renewcommand{\arraystretch}{1.3} 
    \begin{tabularx}\textwidth{|@{}|L|M|c|c|@{}|}
    \hline
    \multicolumn{1}{|c|}{No.} & \multicolumn{1}{|c|}{Equation} & \multicolumn{1}{|c|}{Fit} & \multicolumn{1}{|c|}{Parsimony} \\ 
    \hline
        d1& r(t) =\arcsin(\lambda_{M,4}(t)^2 + \lambda_{M,5}(t)^2) & 26.17 & 14.78\\
       d2& r(t) =1.75 + 1/(1 - \lambda_{M,3}(t)) & 24.91 & 16.0 \\
      d3 & r(t) =1.734731422448 + 1/(1 - \lambda_{M,3}(t)) & 24.11 & 58.30 \\
       d4& r(t) =2.735302247043+(\lambda_{M,5}(t)/((-\lambda_{M,5}(t))+1)) &24.06 & 58.96\\
    \hline
    \end{tabularx}
    \caption{Results for Experiment 2 with polar coordinates}
    \label{tab:exp2_results_polar}
\end{table}
\end{subequations}
\endgroup

\subsection{Experiment 3} \label{sec:exp3}
This subsection considers the experimental setup and implementation in AI Feynman to rediscover the orbital equation of Mars. Knowledge regarding the heliocentricity and planarity of Mars' orbit, discovered from Experiments 1 and 2, are embedded as observational biases into the AI Feynman algorithm. 

\subsubsection{Experimental Setup}
The three-dimensional heliocentric observations of Mars in both the polar and cartesian coordinate systems are reduced to lower dimensional spaces by principal component analysis~\cite{Jolliffe1986}. 

For the cartesian coordinate system, the three-dimensional heliocentric x, y and z coordinates of Mars are projected to a two-dimensional space. The three-dimensional x, y and z coordinates of Mars is $M(t) = [M_x(t), M_y(t), M_z(t)]$. The covariance matrix for Mars is $M(t)^\intercal M(t)$. The eigenvalues and eigenvectors of the covariance matrix are computed. To get a two-dimensional projection of the three-dimensional space, $M(t)$ is projected onto a two-dimensional space defined by the eigenvectors corresponding to the two largest eigenvalues of $M(t)^\intercal M(t)$. 

One run of AI Feynman is made to describe the distance between Mars and the Sun. The inputs are the two projections of Mars in two-dimensional space.

For polar coordinates, the three-dimensional heliocentric polar coordinates of Mars comprise the declination, right ascension and angular width. As the declination and right ascension are angles and the angular width is a measure of length, they cannot be simultaneously projected onto a lower dimensional space. Therefore, only the angles can be projected onto a lower dimensional space. 

The angles of Mars are $M(t) = [M_{RA}(t), M_{DC}(t)]$. The covariance matrix for Mars is $M(t)^\intercal M(t)$. The eigenvalues and eigenvectors of the respective covariance matrices are computed. To get a one-dimensional projection of the two-dimensional space, $M(t)$ is projected onto a one-dimensional space defined by the eigenvectors corresponding to the largest eigenvalue of $M(t)^\intercal M(t)$.

One run of AI Feynman is made to describe the distance between Mars and the Sun. The inputs to the run are the sine and cosine of the projections of the angles in a one-dimensional space.

\subsubsection{Experimental Results}
The equations along the Pareto frontiers are reported in column 2 of Tables~\ref{tab:exp3_results} and \ref{tab:exp3_results_polar} for the cartesian and polar coordinate systems respectively. $\lambda_{M,N}(t)$ denote the projection of Mars with $\lambda_{M,1}(t)$ corresponding to the projection from the eigenvector with the largest eigenvalue. For the cartesian and polar coordinate systems, $N=2$ and $N=1$ respectively. The fit and parsimony of the equations are measured and presented in columns 3 and 4.

\begingroup
\begin{subequations}
\begin{table}[h]
    \tiny
    \centering
    \renewcommand{\arraystretch}{1.3} 
    \begin{tabularx}\textwidth{|@{}|L|M|c|c|@{}|}
    \hline
    \multicolumn{1}{|c|}{No.} & \multicolumn{1}{|c|}{Equation} & \multicolumn{1}{|c|}{Fit} & \multicolumn{1}{|c|}{Parsimony} \\ 
    \hline
        e1 & r(t) = (\lambda_{M,1}^2(t) + \lambda_{M,2}^2(t))^{0.5} &  0.97 & 13.17 \\
        e2 & r(t) = (\lambda_{M,1}^2(t) + \lambda_{M,2}^2(t))^{0.5} + 1.033e-9 & 0.74 &29.83 \\
        e3 & r(t) = (\lambda_{M,2}(t) \times (\lambda_{M,1}(t)^2/\lambda_{M,2}(t) + \lambda_{M,2}(t)))^{0.5} + 1.033e-9 & 0.74 & 36.75 \\
    \hline
    \end{tabularx}
    \caption{Results for Experiment 3 with cartesian coordinates}
    \label{tab:exp3_results}
\end{table}
\end{subequations}
\endgroup  

Equations~\ref{e1}, \ref{e2} and \ref{e3} in Table \ref{tab:exp3_results} suggest that the orbital equation of Mars is a circle, likely because the eccentricity of the orbital equation of Mars is small. An analysis of the projections of Mars reveals that the eigenvectors of the projections correspond to the x- and y-axis of the plane that the two-dimensional ellipse lies on. 

We re-run AI Feynman to rediscover the relationship between the x- and y-coordinates of the planar, two-dimensional ellipse. This run of AI Feynman omits $r(t)$ as an input, and returns the equation $y(t) = (-0.99\times x(t)^2 - 0.28\times x(t) + 2.28)^{0.5}$. This is the cartesian form of Kepler's first law, seen in Equation~\ref{eqn:mars_cartesian}, which suggests an ellipse with a semi-minor axis of 1.5165, a semi-major axis of 1.5241, and an eccentricity of 0.09974. The reported semi-major axis of Mars is 1.5237, and the reported eccentricity of Mars' orbit is 0.09341~\cite{NASA2020MarsFacts}.

\begingroup
\begin{subequations}
\begin{table}[h]
    \tiny
    \centering
    \renewcommand{\arraystretch}{1.3} 
    \begin{tabularx}\textwidth{|@{}|L|M|c|c|@{}|}
    \hline
    \multicolumn{1}{|c|}{No.} & \multicolumn{1}{|c|}{Equation} & \multicolumn{1}{|c|}{Fit} & \multicolumn{1}{|c|}{Parsimony} \\ 
    \hline
        f1 & r(t) = \arccos(0.125\times \cos(\lambda_{M,1}(t))) &24.65 & 8.75  \\
        f2 & r(t) = 1.2 + 0.5/(\cos(\lambda_{M,1}(t)) + 2) & 24.35 & 14.47\\
        f3 & r(t) = 1.201808689874+\frac{\sin(\lambda_{M,1}(t))}{2\times \sin(\lambda_{M,1}(t))\times (\cos(\lambda_{M,1}(t))+2)} & 24.40 & 59.36 \\ 
        f4 & r(t) = 1.041283341747\times \sqrt{((\cos(\lambda_{M,1}(t))/(-(\sin(\lambda_{M,1}(t))+2)))+2)} & 24.22 & 77.73 \\ 
        f5 & \begin{aligned}r(t) &= 1/\left(0.0274130068719387\times \sin(\lambda_{M,1}(t)) + 0.0555427223443985\times \cos(\lambda_{M,1}(t)) \right.\\
        &\left.+ 0.666666666666667\right) \end{aligned}& 22.67 & 99.43 \\
        f6 & \begin{aligned}r(t) &=  1/(0.0274164322763681\times \sin(\lambda_{M,1}(t)) + 0.0555496737360954\times \cos(\lambda_{M,1}(t)) \\
        &+ 0.666666666666667) \end{aligned}& 222.67 & 299.43 \\
        f7 & \begin{aligned}r(t) &=  1/(0.029216912995846\times \sin(\lambda_{M,1}(t)) + 0.0551251780273508\times \cos(\lambda_{M,1}(t)) \\
        &+ 0.666666666666667) \end{aligned}& 22.37 & 99.52 \\
        f8 & \begin{aligned}
            r(t) &=  \arccos \left(0.0584186388859265\times \sin(\lambda_{M,1}(t)) + 0.125\times \cos(\lambda_{M,1}(t)) \right.\\
            &\left.+ 0.0529077895460092\right) 
        \end{aligned}& 22.27  & 100.29\\
        f9 & \begin{aligned}
            r(t) &=  \arccos \left(0.0626782178878784\times \sin(\lambda_{M,1}(t)) + 0.125\times \cos(\lambda_{M,1}(t)) \right.\\
            & \left.+ 0.0570981465280056\right)
        \end{aligned}&  21.84 & 100.50\\
        f10 & \begin{aligned}
            r(t) &= -0.0629538521170616\times \sin(\lambda_{M,1}(t)) - 0.125\times \cos(\lambda_{M,1}(t)) \\
            &+ 1.51335501670837
        \end{aligned} & 21.82 & 101.62 \\
        f11 & \begin{aligned}
            r(t) &=  \arccos \left(0.0626782178878784\times \sin(\lambda_{M,1}(t)) + 0.126976847648621\times \cos(\lambda_{M,1}(t)) \right.\\
            &\left.+ 0.0570981465280056\right)
        \end{aligned} & 21.69 & 140.03\\
        f12 & \begin{aligned}
            r(t) &=  -0.0629538521170616\times \sin(\lambda_{M,1}(t)) - 0.127503842115402\times \cos(\lambda_{M,1}(t)) \\
            &+ 1.51335501670837
        \end{aligned} & 21.61 & 141.16\\
        f13 & \begin{aligned}
            r(t) &= 1/(0.0274130068719387\times \sin(\lambda_{M,1}(t)) + 0.0555427223443985\times \cos(\lambda_{M,1}(t)) \\
            &+ 0.663939893245697)
        \end{aligned} & 20.76 & 142.18 \\
        f14 & \begin{aligned}
            r(t) &= 1/(0.0274164322763681\times \sin(\lambda_{M,1}(t)) + 0.0555496737360954\times \cos(\lambda_{M,1}(t)) \\
            &+ 0.66388201713562)
        \end{aligned} & 20.76 & 142.18 \\
        f15 & \begin{aligned}
            r(t) &= (0.0274164322763681\times \sin(\lambda_{M,1}(t)) + 0.0555496737360954\times \cos(\lambda_{M,1}(t)) \\
            &+ 0.66388201713562)^{-1.00050556659698} \end{aligned}&20.77& 187.69 \\
        f16  &\begin{aligned}
            r(t) &= (0.0274130068719387\times \sin(\lambda_{M,1}(t)) + 0.0555427223443985\times \cos(\lambda_{M,1}(t)) \\
            &+ 0.663939893245697)^{-1.00071978569031}
        \end{aligned}
         & 20.77&  187.69\\
    \hline
    \end{tabularx}
    \caption{Results for Experiment 3 with polar coordinates}
    \label{tab:exp3_results_polar}
\end{table}
\end{subequations}
\endgroup  

Equations~\ref{f5} to \ref{f7} and \ref{f12} to \ref{f16} in Table~\ref{tab:exp3_results_polar} are similar to Equation~\ref{eqn:mars_polar} but have an additional sinusoidal function in the denominator. An analysis of $\lambda_{M,1}(t)$ reveals that the declination and right ascension of Mars have been projected into a one-dimensional space corresponding to an angle between Mars and the Sun with respect to a horizontal, corresponding to the \textit{anomalia coequata}. The sinusoidal function in the denominator expresses the ellipse has a focus that is vertically displaced from the origin. 
We shift the ellipse to remove the vertical displacement and run AI Feynman again. The inputs to the run are the sine and cosine of the projections of the angles in a one-dimensional space. The run uses the inputs to describe the distance between Mars and the Sun. This run of AI Feynman returns the equation $r(t) = \frac{1}{0.0549323\times \cos(\theta(t)) + 0.6612821}$. This is the polar form of Kepler's first law, seen in Equation~\ref{eqn:mars_polar}, which suggests an ellipse with a semi-major axis of 1.5227, and an eccentricity of 0.08306.

\section{Conclusion} \label{sec:conclusion}
In this work, we have expanded the frontier of symbolic regression capabilities by extending AI Feynman using inductive biases to smartly explore changes in reference frames and reductions in dimension spaces. We found heliocentric and two-dimensional, planar projections of the data could parsimoniously fit and describe the distance of Mars from the Sun. This enhancement paves the way for AI Feynman to make paradigm shifts that were previously only possible with human intuition and understanding. Directly embedding an observational bias regarding the discovered heliocentricity and planarity allowed AI Feynman to rediscover the orbital equation of Mars.



\section*{Acknowledgements}
This research is supported by Singapore Ministry of Education, grant MOE-T2EP50120-0019, and by the National Research Foundation, Prime Minister’s Office, Singapore, under its Campus for Research Excellence and Technological Enterprise (CREATE) programme as part of the programme Descartes.

\bibliographystyle{splncs04}
\bibliography{cite}
%




\end{document}